\documentstyle[11pt,newpasp,twoside,epsf]{article}
\def\edcomment#1{\iffalse\marginpar{\raggedright\sl#1\/}\else\relax\fi}
\reversemarginpar

\begin{document}
\title{The Star Formation History of the Milky Way}
 \author{Gerard Gilmore}
\affil{Institute of Astronomy, Madingley Road, Cambridge CB3 0HA, UK}

\begin{abstract}

Quantification of the Galaxy's star formation history involves both
the duration and the rate of formation, with these parameters being
known with different precision for different populations. The early
rate of star formation is knowable from modelling chemical element
data, the recent rate directly from isochrone analyses of
colour-magnitude data. The field halo and globular clusters are almost
exclusively old, and formed in at most a few Gyr. The outer bulge
probably formed in a short period long ago -- extant data is
inconsistent, while the inner bulge/disk is forming today, and has
continued to form over time. Only very limited data is available on
the inner disk. The outer disk near the Sun seems as old as the
halo. The earliest extended disk, which forms the thick
disk today, seems to have been in place very early, an observation
which is not simply consistent with some galaxy formation models.

\end{abstract}

\section{Introduction}

Which is more important:
do galaxies form, or are they assembled? More explicitely, does star
formation occur primarily in the eventual gravitational potential,
albeit in localised regions within that potential, or does it occur in
regions with much smaller potential wells, which are later assembled
in the current whole?

The star formation history of a galaxy, explicitely here our Milky Way
Galaxy, where the most detailed information is attainable, is the
convolution of two functions. One function describes the rate of
formation of the stars which are today in the Galaxy. The second
describes the assembly of those stars into the present Galactic
potential well. There is direct evidence that this assembly continues
today, with both stars and gas being assembled into, or at least
rearranged in, the Galactic potential (eg, the Sagittarius dwarf
spheroidal, Ibata, Gilmore \& Irwin 1994, 1995; the Magellanic
Stream, Putman etal 1998). 
HST imaging suggests that the rate of accretion/merging was
significantly higher in the past, so that accretion has always been
significant. However, HST and other data also show that the star
formation rate was higher in the past.  A topical question is then the
relative importance of these two processes, the normalisation of the
merger and star formation functions, and the order in which they
occur: does star formation mostly happen during mergers,
or after two potential wells have merged?

Some fairly direct constraints on the relative importance of star
formation in a galaxy-scale potential can be deduced from the galaxian
luminosity metallicity relation, shown in figure~1 for the Local
Group. For Local Group galaxies direct metallicities can be
determined, obviating possibly model-dependent interpretation of line
indices for unresolved populations. Nonetheless, this relation is
consistent with the more general mass-metallicity relation for
galaxies, suggesting this relation is universally valid, and
applicable over the full mass range of galaxies.

   \begin{figure}[ht]
\plotone{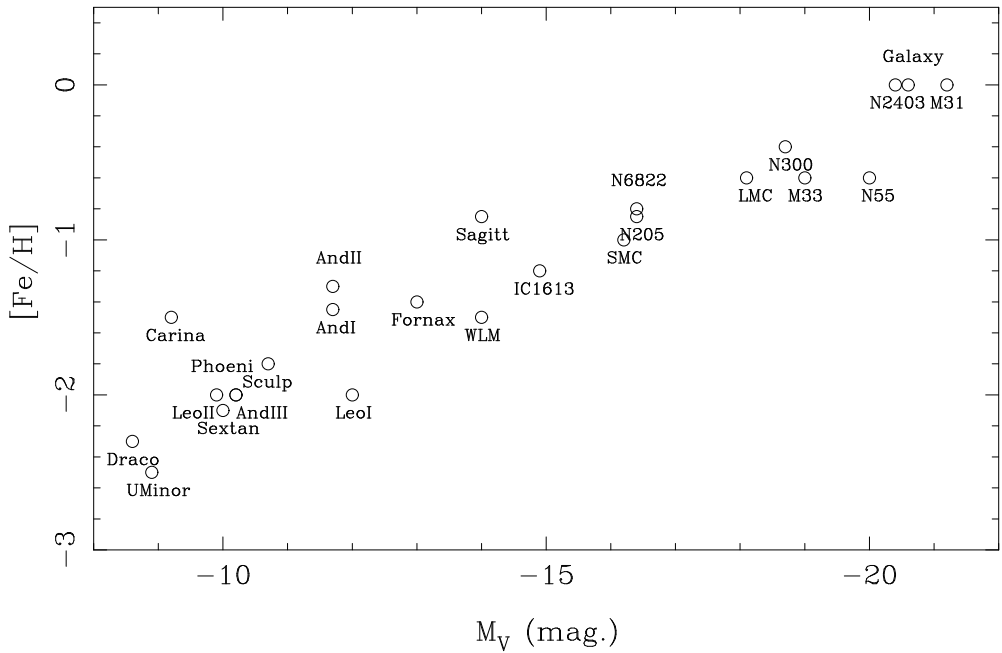}
\caption{The luminosity-metallicity relation for Local group
galaxies. The rather good correlation indicates that the average star
forms in a potential well which is related to the potential well
today, even though the mean stellar age in these galaxies covers a
wide range. Equivalently, one cannot build a large galaxy by
assembling smaller stellar ones.}  \end{figure}

The mass-metallicity relation shown in Fig~1 indicates that when the
typical star in a galaxy is formed, at whatever redshift, it knows the
depth of the gravitational potential in which it will be orbiting at
redshift zero. This knowledge indicates that one cannot form a large
galaxy by assembling smaller stellar systems: any such resultant
object would be of high luminosity and stellar mass, but of low
metallicity. Such objects are not found.  The natural solution is to
form most stars, and most metals, either during mergers, or later
after assembly of gas. Star formation during mergers of comparable
mass gassy systems, while a very visible process which is clearly
important still today, and a natural way to build bulges, is a natural
way to build a disk galaxy only if star formation is very inefficient
{\sl during} the merger. It is also far from a natural way to build a
very low-mass galaxy.  One might be tempted to conclude that low mass
galaxies and galaxy disks formed their stars not only {\sl in situ},
but at a lower rate than was relevant to bulge formation.  We now
consider local constraints on this speculation.

\subsection{ Global considerations}

The rate at which stars formed, on average in tolerably luminous
systems, has been recently quantified in the Madau plot. In linear
time, rather than redshift space, this shows a roughly constant rate
of star formation from very early times ($\la 1$Gyr afer time zero)
continuing for 4$-$5Gyr, then an apparently rapid decline (no doubt
exaggerated by the non-linear time-redshift relation) by a factor of
about 4$-$5 to a new roughly constant rate, continung for a further
5$-$6 Gyr until today. Thus, some two-thirds of stars were formed
before redshift about unity. Does the Local Group follow this trend?
If it does, then both the whole of their bulges, and the inner disks
of the Galaxy and M31 must have been in place, and stellar, before
redshift unity.

We discuss this further below, but note here that the extant, albeit
indirect evidence, suggests consistency with the Madau plot.  The most
directly consistent interpretation is that the bulge and early disk
(now the thick disk) were formed in the first $\sim 2-3$Gyr, at
redshifts significantly greater than unity.

\section{ Star Formation Histories: Indirect methods}

A natural calibration of  rapid rates of star formation is available from
the dependence of the creation sites for some chemical elements on the
main-sequence mass, and hence life-time, of the pre-supernova
star. The most important elements in this regards are the
$\alpha-$elements, especially oxygen, calcium, silicon and magnesium.
These $\alpha-$elements are created and expelled during the type~II
supernovae of stars with initial masses in excess of about 10M$_\odot$,
and so become available to enrich newly forming stars on times of
$10^8$yr after the initiation of significant star formation. Iron-peak
elements are primarily created and expelled in the type~I supernova of
lower mass stars, with characteristic times of $\ga 10^9$ years. 

Thus, a high relative abundance of the $\alpha-$elements, compared to
the iron-peak elements, indicates that the corresponding star formed
within $\la 10^9$years of the onset of significant star formation.
Correspondingly, if most stars in a `population' are determined to
have relatively high values of the $\alpha-$elements then one may
deduce that those stars all formed within one Gyr of the onset of
significant star formation.

This situation, together with some recent observational data, is
summarised in Figure~2. The left hand panel provides results from a
recent study of 90 disk F-and G-dwarfs (Chen, Nissen, Zhao, Zhang \&
Benoni 2000). They, consistently with all other studies (eg Fuhrmann
1998), show the systematic overabundance of the $\alpha-$elements at
metallicities below about $-0.5$dex. The right hand panel provides the
complementary models, and other data on more metal poor stars. This
figure, taken from Gilmore \& Wyse (1998), shows the location of the
metal-poor field halo stars as the hatched region, similar data from
the literature to that in the left panel on thick disk stars, and a
set of simple models.

\bigskip
\begin{figure}[h!tb]
\plottwo{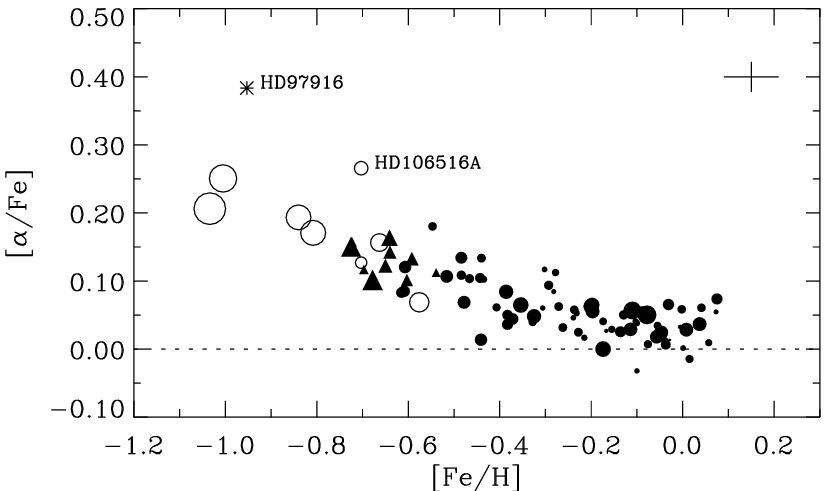}{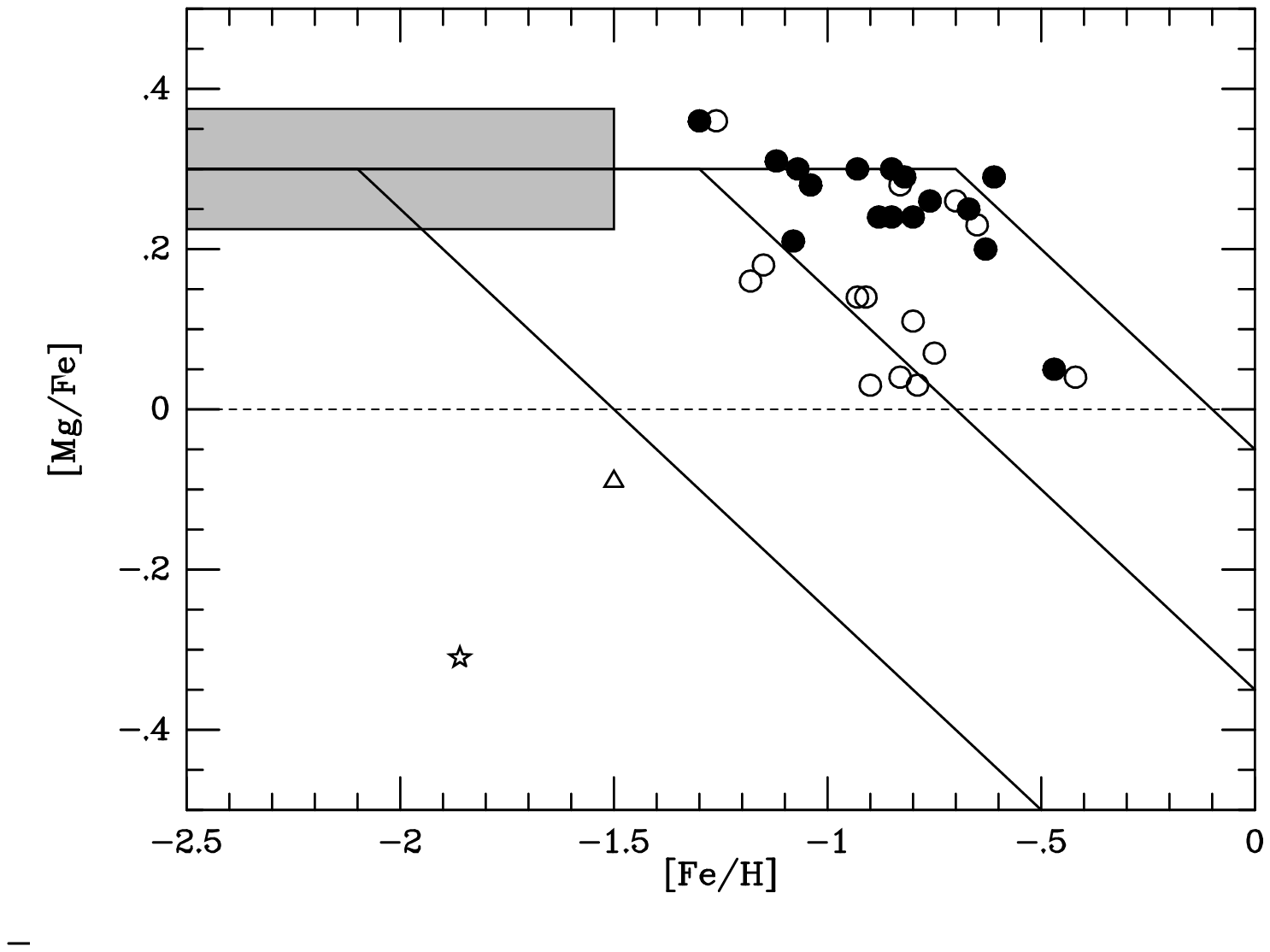}
\caption{Left panel: $\alpha-$element overabundance as a function of
iron-peak abundance for a sample of old disk stars, from Chen etal
(2000). The systematic overabundance of the $\alpha-$elements through
the thick-disk region, below about -0.5dex, is direct evidence for a
rapid formation of all the stars in this abundance range. Right panel:
(from Gilmore \& Wyse 1998). The shaded area locates the mean
metal-poor halo field star, while the sloping lines illustrate models
of star formation extending over many Gyr. These simple models confirm
that the thick disk formed most of its stars in an interval of at most
a few Gyrs after the initiation of star formation.}

\end{figure}

The crucial conclusion, immediately apparent from the data, is that
the thick disk stars with metallicity between $-$1dex and $-$0.5dex have
the same $\alpha-$element overabundance as does the metal-poor halo,
with rather few exceptions. That is, both the field halo and the thick
disk formed within $1-$2Gyr of the onset of star formation in their
respective locations. We emphasise that this does not imply any
relationship between the halo and thick disk populations: it simply
means that both formed rapidly, once they started to form. Absolute
age dating the populations is a separate problem, which we consider
below.

We note in passing that the normalisation of the $\alpha-$element
enhancement depends on the slope of the stellar IMF above the SNII
mass limit, about 10M$_\odot$. The similarity between this
normalisation for the stars with metallicty near $-$0.5dex and those
near $-$2.5dex is strong evidence for an invariant high-mass IMF slope
across this metallicity range.

Observational data appropriate for analyses of this type have rapidly
increased in quality and quantity in the recent past, with very many
studies now available. Unfortunately as yet, inadequate data is
available to extend this analysis to the Galactic bulge: published
abundance data for bulge stars are inconsistent with stellar
production ratios derived from supernova observation and models, and
so do not allow a self-consistent analysis. The availability of
VLT+UVES is expected to remedy this lack in the near future, and is an
exciting prospect, allowing robust determination of the formation rate
of a galactic bulge for the first time.

\section{Direct Age Determinations}

The most robust absolute age determinations of course involve
isochrones, but are feasible only when reliable distances
and accurate photometry for individual stars are available. This
applies reasonably well to special cases, such as globular and open
clusters, and satellite galaxies, and the Solar neighbourhood. Age
limits may also be derived without distances when a population has an
approximately known abundance range, and is predominantly old.

\subsection{Satellite galaxies}

The star formation histories of the existing/surviving low surface
brightness dwarf companions to the Milky Way are varied, with all star
formation histories being apparent 
(eg Mateo 1998). Recent advances in variational calculus inversion
methods have provided objective star formation histories of the
satellite dSph galaxies (Hernandez, Valls-Gabaud \& Gilmore 2000a, and
refs therein), showing that the star formation history, averaged over
the sample, is crudely constant with time. The absolute rate is
additionally determined with this method, and is extremely low:
unsurprisingly, given the shallow depth of the corresponding potential
wells. 

\subsection{The Galactic Bulge and Inner Disk}

HST studies of the outer Galactic Bulge, while remaining fraught with
complexity, are consistently showing that at least the bulge more than
a few COBE scale lengths from its center is old (Ortolani etal 1995;
Feltzing \& Gilmore 2000). The detailed recent work has confirmed an
assumption in earlier analyses, that, at least statistically, the very
many young stars seen in the line of sight towards the Bulge are
distributed spatially like the inner disk, rather than like the bulge.

The central kpc of the Plane of course is one of the highest star
formation rate parts of the Galaxy, and one can but wonder where the
middle-aged descendents of such stars which formed in the past are to
be found today. The current star formation rate in the central galaxy
is sufficient to build the bulge over a Hubble time, yet where are the
intermediate age stars? This highlights our near complete ignorance of
the age distribution in the inner disk.

Recent ISO survey data, and its spectroscopic follow-up,
is identifying a substantial population of intermediate age AGB stars
in the inner Galaxy, confirming continuing star formation over time in
the inner disk (Omont et al 1999; van Loon, Gilmore, etal, in preparation). 
Any plausible age-velocity dispersion relation should have scattered
these stars into what is classically called `the bulge', given the
similarity of scale thicknesses of the inner disk and the COBE
bulge. Yet they remain to be identified. Perhaps the most metal-rich
gK stars, for which ages are not yet available, are in
fact substantially younger than the more metal-poor stars.

\subsection{Globular Clusters}

Impressive recent studies have shown the majority of
globular clusters in the halo are  old, with a remarkably small age
spread (e.g. Rosenberg et al.~1999), while there is a small subset,
particularly among the more metal-rich clusters, with inferred ages of
several Gyr younger than the dominant old population. 

\subsection{Field Population II}

Age limits are available for stars in the field halo, for stars whose
orbits probe most of available phase space.
Figure~3, for a kinematically-selected, local sample, which through
the orbits of the stars, probes a significant part of the stellar
halo, shows that the vast majority of field halo stars are old,
but there is a small fraction, as for the globular clusters
biased to the more metal-rich
stars, that are candidates for being several Gyr younger.  Normalizing
through the local halo metallicity distribution, and interpreting
generously all stars blueward of the old turnoffs as being truely
younger, implies that at most only around 10\% of the stellar halo
could be `intermediate-age' (Unavane, Wyse \& Gilmore 1996). 

\begin{figure}[!h]
\plotfiddle{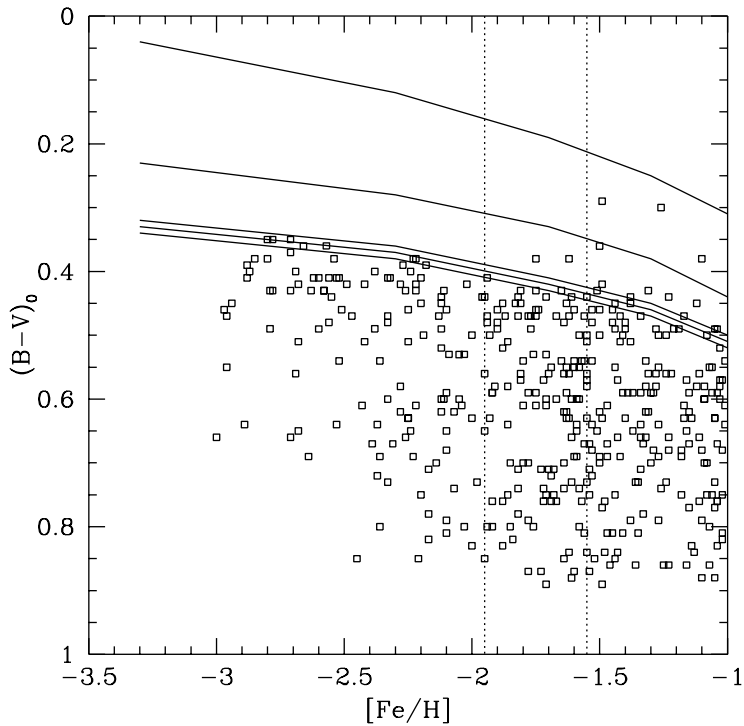}{3.5in}{0}{50}{50}{-150}{-80}
\caption{Scatter plot of de-reddened B-V colour against metallicity
for the local, kinematically-selected, unevolved halo stars in the
Carney et al.~(1994) sample.  Uncertainties have been ignored in the
interests of clarity, and are of order 0.1 dex in [Fe/H] and 0.01 in
colour.  Superposed lines indicate the main sequence turn-off colour
as a function of metallicity for isochrones of age
\{8,10,15,16,17\}Gyr (using the Revised Yale Isochrones).  Stars with
a {\it bluer\/} colour than a given turnoff, at fixed metallicity, are
candidates for being younger than that isochrone. (Taken from Unavane,
Wyse \& Gilmore 1996.)}
\end{figure}

This is in  agreement with the fraction of anomalously blue
halo stars found through very different selection criteria by Preston,
Beers \& Schectman (1994).  However, Preston \& Sneden (2000) have
analysed the chemical compositions and possible radial velocity
variations of 62 of the Preston et al.~(1994) `Blue Metal-Poor' stars,
for which an intermediate-age had been ascribed.  They find that a
very large fraction of these stars are in binaries, indeed with binary
parameters suggestive of mass transfer as the explanation for their
colours, rather than relative youth.  The possible intermediate-age
fraction of the halo is then reduced by at least a factor of two below
the earlier 10\% estimate (Preston \& Sneden 2000). 

\subsection{The Galactic Disk}

The star formation history of the Galactic disk is of interest not
only {\sl per se}, but also because it provides the most direct test
of galaxy formation/merger  models. Current understanding of the disk
considers two recognizable phases: an early disk, later dynamically
heated, probably by the last significant Galactic merger, creating
what we now call the thick disk; and a later/continuing thin disk, forming
undisturbed at a low rate.

The chemical element ratio evidence that the whole of the thick disk
formed within 1$-$2Gyr of the onset of its star formation is reviewed
above. Absolute age dating of the thick disk remains somewhat
problematic: it is difficult to isolate an unambiguous sample of thick
disk stars. In so far as this has been attempted, however, the
analysis suggests no detectable age interval between formation of the halo and
bulge, and formation of the thick disk (eg Binney, Dehnen \& Bertelli
2000; Mendez \& Ruiz 2000). Even interpreting this `no age interval'
conservatively still implies that the Milky Way had a disk, with scale
length of about 3kpc, and already having as stars some ten percent of
today's disk luminosity, at redshifts of 2 or thereabouts. A
challenge for some models.

The more recent evolution of the disk, and a methodology which will
eventually quantify the whole situation, when adequate data are
available, is summarised in figure~4. This figure, from Hernandez,
Valls-Gabaud \& Gilmore (2000b), shows the Solar neighbourhood star
formation history, derived from Hipparcos data, using a non-parametric
inversion method. While the time baseline here is severely restricted
by the Hipparcos sample, data of higher quality and allowing extension
of such analyses to the Galactic inner bulge will be available from
GAIA (Gilmore et al 2000).

\begin{figure}[h]
\plottwo{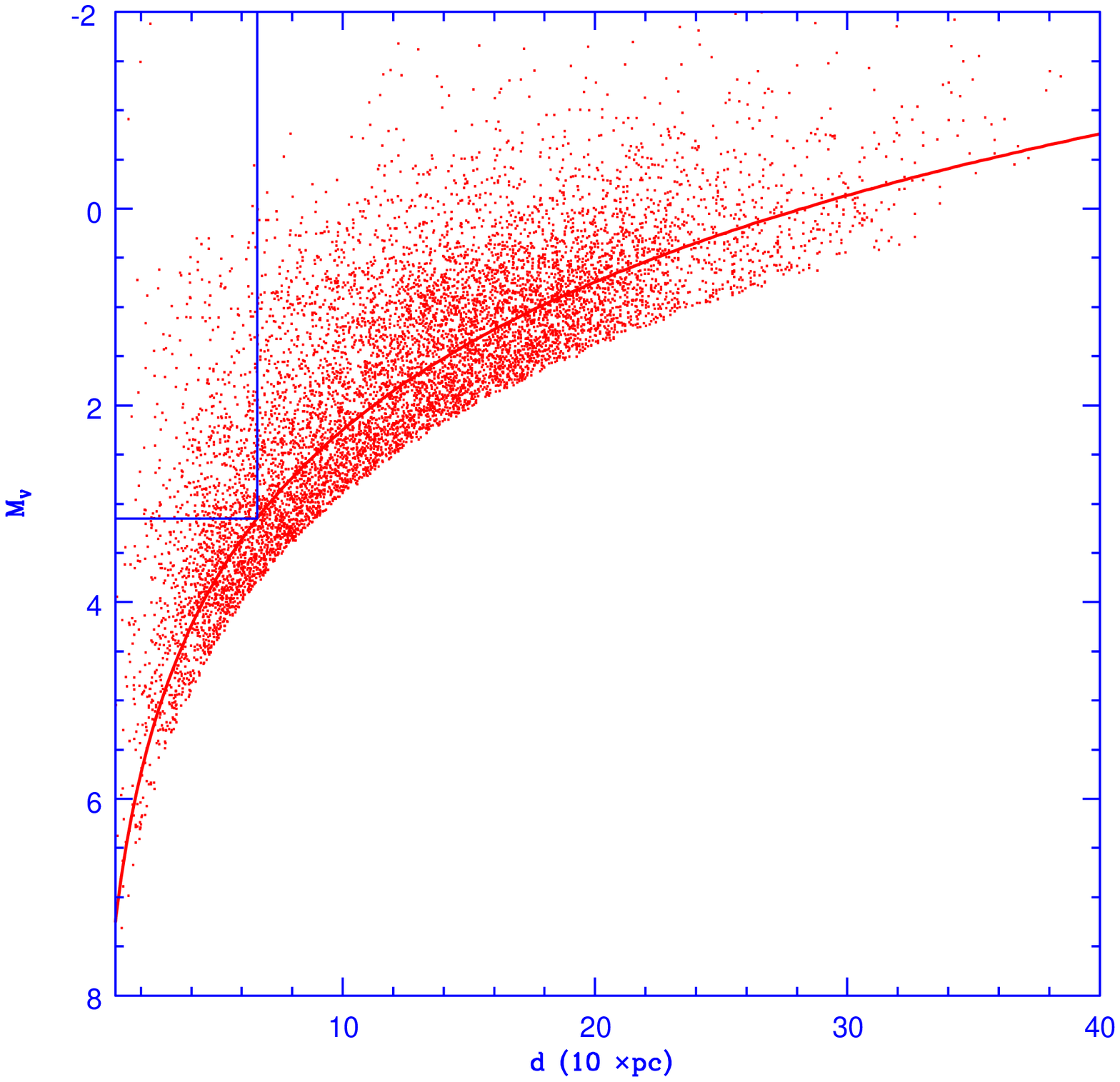}{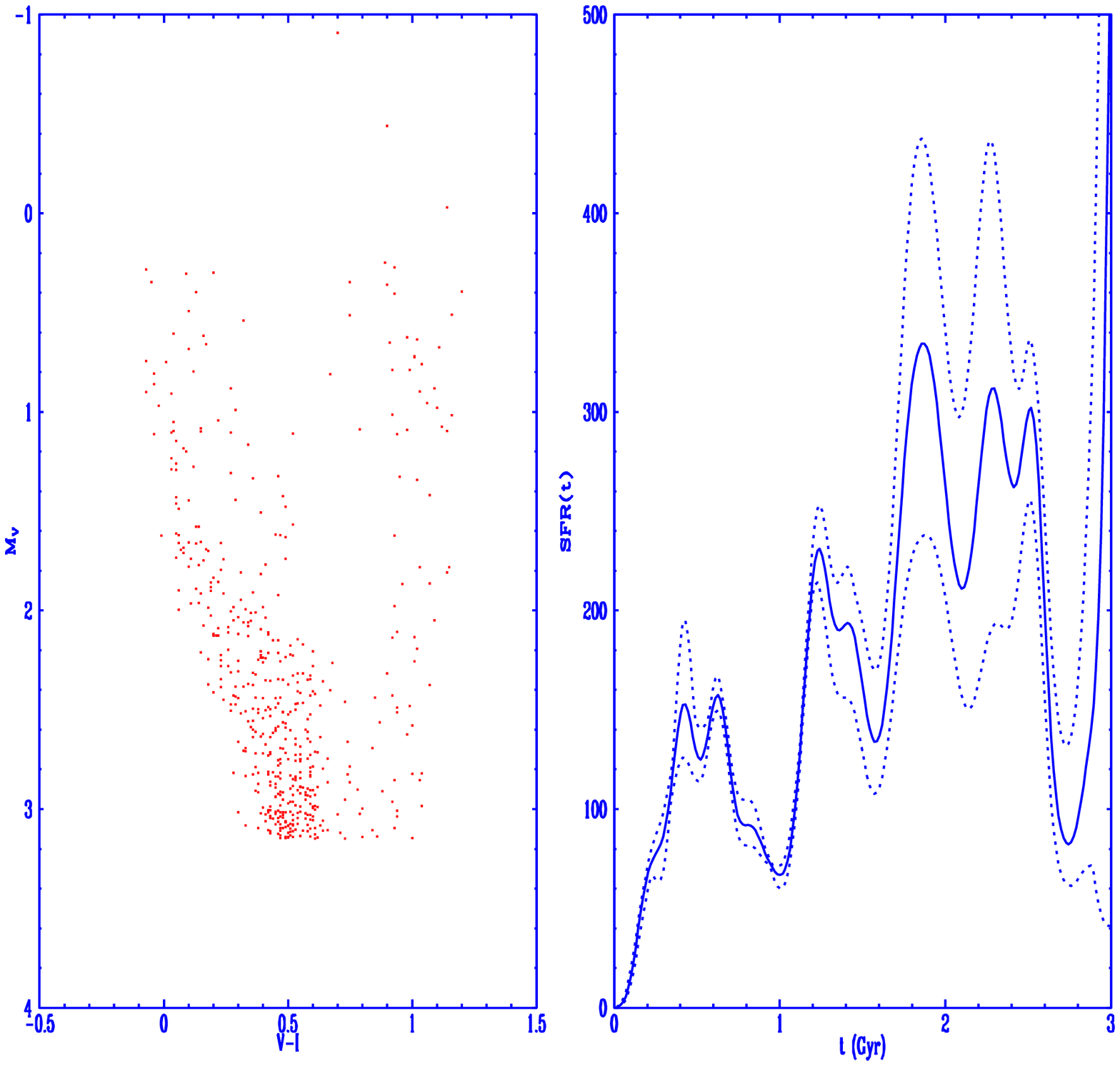}
\caption{Left Panel: the Hipparcos sample of stars in absolute
magnitude-distance space, from which the local star formation history
can be derived. Right panel: the resulting star formation history of
the Solar Neighbourhood, over the last 3Gyr, derived from the
Hipparcos data using a non-parametric variational calculus
technique. The decline in the last 0.5Gyr is an artefact of the Sun's
location: the earlier result is robust. These figures are taken from
Hernandez, Valls Gabaud \& Gilmore 2000b, where details of the method
can be found. }
\end{figure}

\section{ IMPLICATIONS for STAR FORMATION RATES}

One may quantify the general discussions above, combining limits on the
chemical uniformity of a stellar population, the element ratio
information and its scatter, and the luminosity/number of stars
involved, to derive limits on actual star formation rates.

The first limit comes from the observed small scatter in the ratio of
the $\alpha-$elements to the iron-peak elements at a specific
metallicity. This small scatter requires one of two orthogonal
conditions to be met: either the duration of star formation was so
short that no self-enrichment took place, a situation naturally
consistent with no range in any elemental abundances (globular cluster
formation?), or the duration of star formation was sufficiently long,
and the rate sufficiently low, that efficient mixing of SNae ejecta
across the whole star-forming volume was possible. To minimise
variations due to shot-noise in the number of SNae `enrichment
events', the involved volume cannot have been too small.

To quantify the duration and rate {$\cal R$} of star formation in the
stellar populations of the Galactic bulge and halo we need to specify
four observational quantities: the halo (or bulge) stellar mass
($2.10^9 M_{\odot}$), the lifetime and 
mass of the stellar supernova progenitors (available from
isochrones: $\sim 10M_{\odot}$, $\tau \approx2.10^7yr$),
the mass fraction of the halo which has been sampled by extant
spectroscopic element ratio data ($\ga 50\%$), and the fraction
$\eta$ of halo stars which deviate significantly from the predominant
halo element ratio distribution ($\eta \sim 10\%$). 

In addition, we must adopt one model-dependent number, a mixing
efficiency term $\zeta$. With this situation the star 
formation rate {$\cal R$} limitation becomes:

\begin{equation}
{\cal R} \la \zeta \/ \eta \/ \tau^{-1} \/ M_{halo}
\end{equation}

The mixing efficiency parameter can be quantified from simple Monte
Carlo simulations of the number of supernovae ejecta, with each SN
event drawn randomly from the available stellar IMF mass range, 
$10M_{\odot} \la M_{SN} \la 100M_{\odot}$, and assigning to each
the appropriate (theoretical) mass-dependent yield. The number of
SNae, and the mass of gas into which the ejecta are mixed, which are
required to keep an observed scatter below that seen in stars can then
be determined. The minimum requirement is for $\ga$ 30 SNae and a gas
mass $\ga 10^5 M_{\odot}$ in a well-mixed region.

These parameters may be converted into a corresponding length scale,
and a consequent limit on the mixing efficiency,
adopting a length scale relevant to observations at high redshift. For
the present, we adopt the relatively high densities of damped
Ly$-\alpha$ systems, implying length scales of tens of pc for n$_{\rm
H} \sim 10^{21}$, and corresponding sound speeds of v$_c \sim 1$km/s,
for the smallest possible well-mixed regions. 

Assuming all halo star formation takes place in a set of regions of
this small size provides a
conservative limit on the mixing efficiency parameter $\zeta$ of order ten.
More uniformly distributed star formation would require even lower
star formation rates to allow adequate mixing of the SNae ejecta.

From this we deduce the halo star formation rate required to be
consistent with the scatter observed in field stars today of:
\begin{equation}
{\cal R} \la 10 ({\eta\over{0.1}}) ({M_{halo}\over{2.10^9 M_{\odot}}})
({{2.10^7 yr}\over{\tau}}) \/\/ M_{\odot} yr^{-1}.
\end{equation}

The deduction is therefore that field halo star formation lasted $\ga
2.10^7$ yr, and at a mean rate of ${\cal R} \la 10M_{\odot}
yr^{-1}$. Similar arguments apply to the bulge.

\begin{table}[h]
\begin{center}
\caption{A summary of star formation rates, and durations of star
formation, in some Galactic stellar populations. These values are
derived from combination of chemical element scatter and masses.}
\vskip 10pt
    \leavevmode    
\Large
    \begin{tabular}[tbh]{|l|c|c|}
\hline && \\[-4pt]
{ Population }  & Duration & Formation Rate \\
\hline && \\[-4pt]
  &  (years) & ${\mathcal M}_{\odot}yr^{-1}$ \\
\hline && \\[-4pt]
Globular cluster & $\le10^8$ & $\ge 0.01$ \\
$\omega$Cen&$\ge10^8$ & $\le 0.1$ \\
Halo, [Fe/H]$\le-2.0$ & $\le10^8$ & $\sim 1$ \\
Halo, [Fe/H]$\sim-1.5$ & $\le10^9$ &  $\sim 1$ \\
Bulge; high [$\alpha$/Fe] & few.10$^8$ & 10-100 \\
Bulge; low [$\alpha$/Fe] & few.10$^9$ & 10-100 \\
Thick Disk & few.10$^9$ & 1-10 \\
Current Disk & $10^{10}$ & $\sim 1-10$ \\
Inner Disk & ? & ? \\
Satellite dSph & many.10$^9$ & $\le 10^{-3}$ \\
\hline && \\[-4pt]
 Assembly & early & \\
\hline && \\[-4pt]
Infall & continuing? & $\sim 4$Gyr \\
\hline
\end{tabular}
\vspace{-0.5cm}
\end{center}
\end{table}

In neither case is
there evidence for a very high star formation rate `SCUBA-source'
history. Fundamentally, there are just too few stars in the halo, or
the bulge, for a high star formation rate to have been involved in
their formation, given the lower limit on the duration of star
formation required by the lack of element ratio scatter.

Similar considerations apply to the formation of the thick disk, where
a similar overabundance of the rapidly-formed $\alpha-$elements is
seen. 

For those globular clusters where no evidence of self-enrichment is seen
in the (delta-function) distribution of iron-peak elements, a lower
limit on star formation can be deduced, simply by requiring that all
observed stars are formed before the first SNae occur. This limit is
however very low, of order $0.1M_{\odot} yr^{-1}$. In cases such as
$\omega$Cen, where an abundance spread is seen, a corresponding upper
limit is available, in that the cluster stars should not have all
formed before the self-enrichment could occur. 

These limits on past star formation rates are summarised in Table~1,
which forms the Conclusions of this paper.

\end{document}